\documentclass[12pt,reqno]{amsart}
\usepackage{amsmath,amssymb,amsfonts,amsthm}
\usepackage[mathscr]{eucal}
\usepackage[all]{xy}
\usepackage{hyperref}
\usepackage{setspace}
\allowdisplaybreaks

\begin{document}

\title[Variational principle for cylindrical curves and dynamics of spinning particles]
{Variational principle for cylindrical curves and dynamics of
spinning particles in $d=3$ Minkowski space}

\author{D. S. Kaparulin, S. L. Lyakhovich, I. A. Retuntsev} 
 \email{dsc@phys.tsu.ru, sll@phys.tsu.ru, retuntsev.i@phys.tsu.ru}

\address{ Physics Faculty, Tomsk State University, Tomsk 634050,
Russia }


\begin{abstract} We proceed from the fact that the classical paths of irreducible
massive spinning particle lie on a circular cylinder with the
time-like axis in Minkowski space. Assuming that all the classical
paths on the cylinder are gauge-equivalent, we derive the equations
of motion for the cylindrical curves. These equations are
non-Lagrangian, but they admit interpretation in terms of the
conditional extremum problem for a certain length functional in the
class of paths subjected to the constant separation conditions. The
unconditional variational principle is obtained after inclusion of
constant separation conditions with the Lagrange multipliers into
the action. We explicitly verify that the states of the obtained
model lie on the co-orbit of the Poincare group. The relationship
with the previously known theory is demonstrated.
\end{abstract}

\maketitle

\section{INTRODUCTION}

The classical spinning particles are studied for decades starting
from the paper of Frenkel \cite{Fren}. The first seventy years of
the development and corresponding bibliography are briefly
summarized in \cite{Fryd}. The discussion of recent researches and
references can be found in \cite{Kassandrov}, \cite{Rempel}.

The Kirillov-Kostant-Souriau \cite{KIrillov}, \cite{Kostant},
\cite{Souriau} geometric quantization method tells us that the
quantization of the classical spinning particle model corresponds to
the irreducible representation of the Poincare group if the
classical limit of the dynamical system is the co-adjoint orbit. The
classical action functional is defined by the symplectic form on the
co-orbit. The latter fact means that the momentum and total angular
momentum are the only independent gauge invariant observables in the
irreducible spinning particle theory. The admissible values of these
quantities are restricted by the mass-shell and spin-shell
constraints.

The irreducible spinning particle models are known in
four-dimensional Minkowski and (A)dS spaces \cite{LSS},
\cite{Staruszkiewicz}, \cite{Bratek}, \cite{KLSS}. For lower and
higher dimensions we refer the articles \cite{LSSh1}, \cite{LSSh2},
\cite{LSSh3}. In all these models the physical degrees of freedom
involve besides particle's position in space-time, the position in
the "internal space". The structure of internal space depends on the
model. In particular, the spinor, twistor, vector, and tensor
spinning particle models are distinguished \cite{Fryd}. The common
feature shared by the irreducible particle models is that the
classical paths of free particle lie on a certain surface in
space-time irrespectively to the specifics of the model.

The geometry of irreducible classical paths has been studied in
\cite{LK_WS}. In this work, it has been shown that the gauge
equivalence class of particle's paths necessarily forms a
cylindrical surface in Minkowski space called the world sheet. In
$d=3,4$, the world sheets are $2d$ circular cylinders with a
time-like axis. The cylinder position is determined by the particle
momentum and total angular momentum, while the representation
defines the cylinder radius. The geometrical equations describing
the motion of the particle along the world sheet are obtained for
space-time dimension $d=3,4$ in the papers \cite{LK_WS}, \cite{KLR}.
These equations involve higher derivatives, and they are
non-Lagrangian.

In the present paper, we construct the action functional for a
massive spinning particle in $3d$ Minkowski space from the
requirement that its paths are general cylindrical curves. We
proceed as follows. First, the auxiliary variables are introduced to
reduce the order of cylindrical curve equations. The obtained system
of equations is non-Lagrangian, but it admits interpretation in
terms of conditional extremum problem in the class of constant
separation curves \cite{Starostin}. The non-variational dynamics is
embedded into the variational one after inclusion of constant
separation conditions with the Lagrange multipliers into the action.
We explicitly verify the obtained action describes irreducible
particle. At the final step, we establish the relationship with the
previously known spinning model from the work \cite{GKL}.

\section{CIRCULAR CYLINDERS AND CYLINDRICAL CURVES}
\par
Consider $3d$ Minkowski space with the local coordinates $x^\mu\,,\,
\mu=0,1,2$ equipped with the metric with the mostly positive
signature,
\begin{equation}\label{metric}
       \phantom{\frac12}\eta_{\mu\nu}=\text{diag}(-1,1,1)\,,\qquad \mu,\nu=0,1,2\,.\phantom{\frac12}
\end{equation}
Introduce the circular cylinder with time-like axis as the
hypersurface defined by the equation
\begin{equation}\label{cylinder}
       \phantom{\frac12}(x-a)^2+(n,x)^2=r^2\,.\phantom{\frac12}
\end{equation}
Here, the round brackets denote the scalar products,
\begin{equation}\label{}
    \phantom{\frac12}(n,x)=n_\mu x^\mu,\qquad
    (x-a)^2=(x-a)_\mu(x-a)^\mu\,.\phantom{\frac12}
\end{equation}
All the tensor indices are raised and lowered by the metric. The
real number $r$ is the cylinder radius. The parameters $n$, $a$
determine the cylinder position in space, and they meet the
conditions
\begin{equation}\label{}
    \phantom{\frac12}(n,n)=-1\,,\qquad (n,a)=0\,.\phantom{\frac12}
\end{equation}
The geometric meaning of $n,a$ is follows: $n$ is the normalized
tangent vector to the cylinder axis, and $a$ connects the cylinder
axis and the origin by the shortest path.

A cylindrical curve $x(\tau)$, with $\tau$ being the parameter on
them, is a curve traced on a cylinder (\ref{cylinder}). The
cylindrical curves are known to be the paths of irreducible massive
spinning particles. Once the classical trajectory of the particle
lies on the cylinder (\ref{cylinder}), the linear momentum $p$ and
total angular momentum $J$ of the particle are determined by the
formula from the paper \cite{LK_WS},
\begin{equation}\label{pJ-na}
    \phantom{\frac12}p=mn,\qquad J=m[a,n]-sn\,,\phantom{\frac12}
\end{equation}
where $m,s$ are particle's mass and spin. The square bracket denotes
the cross product of two vectors,
\begin{equation}\label{}
    \phantom{\frac12}[n,a]_\mu=\varepsilon_{\mu\nu\rho}n^\nu a^\rho\,,\phantom{\frac12}
\end{equation}
with $\varepsilon_{\mu\nu\rho}$ being the $3d$ Levi-Civita symbol.
We use the convention $\varepsilon_{012}=1$ throughout the article.
The quantities $p,J$ automatically satisfy the mass-shell and
spin-shell constraints,
\begin{equation}\label{mass-spin-shell}
    \phantom{\frac12}p^2=-m^2\,,\qquad (p,J)=ms\,.\phantom{\frac12}
\end{equation}
This ensures the irreducibility of the particle moving the
cylindrical path.

We want to describe the cylindrical curves by the ordinary
differential equations. In our classification of cylindrical paths,
we mostly follow \cite{Starostin}. We say that the curves $x(\tau)$,
$y(\tau)$ have constant separation $r$ if
\begin{equation}\label{const-sep}
       \phantom{\frac12}(\dot{x},x-y)=0\,,\qquad (x-y)^2=r^2\,.\phantom{\frac12}
\end{equation}
Here, the dot denotes the derivative by $\tau$. Conditions
(\ref{const-sep}) imply that $x(\tau)$ and $y(\tau)$ have a common
normal of fixed length $r$ at any moment of time. The cylinder path
$x(\tau)$ and its projection onto the cylinder axis $y(\tau)$ are
particular examples of constant separation curves. Whenever
$x(\tau)$ runs over the cylinder, the curve $y(\tau)$ follows a
rectilinear path. So, we can select the cylindrical paths by
imposing the differential equation on the curve $y(\tau)$:
\begin{equation}\label{dyddy}
    \phantom{\frac12}[\,\dot{y}\,,\ddot{y}\,]=0\,.\phantom{\frac12}
\end{equation}
As it has been proven in [19], the differential equations
(\ref{const-sep}), (\ref{dyddy}) are satisfied if and only if the
curve $x(\tau)$ lies on a cylinder. The formal proof is quite long,
and we do not repeat it here.

Let us now consider another form of equations (\ref{const-sep}),
(\ref{dyddy}). Introduce the difference vector $d$ that connects the
current particle position on the cylinder and cylinder axis by the
shortest path,
\begin{equation}\label{d-def}
    \phantom{\frac12} y=x-d\,.\phantom{\frac12}
\end{equation}
In terms of the vector $d$ and current particle's position $x$, the
conditions of constant separation (\ref{const-sep}) read
\begin{equation}\label{d-cyl}
    \phantom{\frac12}(\dot{x},d)=0\,,\qquad d^2-r^2=0\,.\phantom{\frac12}
\end{equation}
The substitution (\ref{d-def}) brings the differential equations
(\ref{dyddy}) to the following form:
\begin{equation}\label{dxddx}
    \phantom{\frac12}[\,\dot{x}-\dot{d}\,,\ddot{x}-\ddot{d}\,]=0\,.\phantom{\frac12}
\end{equation}
We mostly work with equations (\ref{d-cyl}), (\ref{dxddx}) because
the dynamical variables $x$, $d$ have clear geometrical
interpretation.

\section{VARIATIONAL PRINCIPLE}
\par

In this section, we derive the variational principle for the
equations (\ref{d-cyl}), (\ref{dxddx}) of cylindrical curves.

We begin form geometric meaning of equations (\ref{d-cyl}),
(\ref{dxddx}). The condition (\ref{dxddx}) ensures that the
projection of the cylindrical path on the cylinder axis is a
straight line. The constant separation condition (\ref{d-cyl})
selects the set of possible particle's positions with one and the
same projection on the axis. Being considered in itself, equation
(\ref{dxddx}) comes from the least action principle for the
functional
\begin{equation}\label{S0-0}
    L[x(\tau),d(\tau)]=-m \int \sqrt{-(\dot{x}-\dot{d\,\,})^2} \, d\tau\,,
\end{equation}
while (\ref{d-cyl}) are additional conditions. Here, the parameter
$m$ with the mass dimension can be interpreted as the particle's
mass, but there is no room for spin. To introduce spin, we consider
the more general series of action functionals parameterized by the
parameter $s$,
\begin{equation}\label{S0}
    S_0[x(\tau),d(\tau)]=-m \int \sqrt{-(\dot{x}-\dot{d\,\,}
    +\frac{s}{mr^2}[\,d,\dot{d}\,])^2} \,
    d\tau\,.
\end{equation}
With account additional conditions (\ref{d-cyl}), the extremals of
the action are cylindrical curves. The sign under the root is chosen
from the fact that the cylinder axis has a time-like tangent vector.

The specifics of the variational problem (\ref{S0}) is that the
extremum of action is sought in the class paths subjected to the
constant separation conditions (\ref{d-cyl}). This setting is
somewhat similar to the optimal control problem \cite{Agrachev},
where the certain functional is extremized in the class of
trajectories that meet differential equations. To construct
unconditional variational principle, the constraints (\ref{d-cyl})
and the Lagrange multipliers $\alpha,\beta$ are added to the action
(\ref{S0}),
\begin{equation}\label{S}
       S[x(\tau),d(\tau),\alpha(\tau),\beta(\tau)]=S_0[x(\tau),d(\tau)]+\int
       \alpha(\dot{x},d)+\frac{\beta}{2}(d^2-r^2)\, d\tau\,.
\end{equation}
In principle, the inclusion of Lagrange multipliers with constraints
into the action can lead to appearance of parasitic degrees of
freedom. This does not happen in this model.

Let us see the equations (\ref{d-cyl}), (\ref{dxddx}) come from the
least action principle for the functional (\ref{S}). The
Euler-Lagrange equations read
\begin{equation}\label{eom1}
       \frac{\delta S}{\delta
       x}=-\frac{m}{(-(v,v))^{\frac{1}{2}}}\dot{v}-\frac{m(v,\dot{v})}{(-(v,v))^{\frac{3}{2}}}v-\alpha\dot{d\,\,}-\dot{\alpha}d=0\,,
\end{equation}
\begin{equation}\label{eom2}
       \frac{\delta S}{\delta d}=\alpha\dot{x} +\beta
       d+\frac{m}{(-(v,v))^{\frac{1}{2}}}\dot{v}+\frac{m(v,\dot{v})}{(-(v,v))^{\frac{3}{2}}}v
       -\frac{s}{r^2}\frac{(v,\dot{v})}{(-(v,v))^{\frac{3}{2}}}[v,d]
       -\frac{s}{r^2}\frac{[\dot{v},d]+2[\,v,\dot{d\,\,}]}{(-(v,v))^{\frac{1}{2}}}=0\,,
\end{equation}
\begin{equation}\label{eom3}
       \phantom{\sum^1}\frac{\delta S}{\delta \alpha}=(\dot{x},d)=0\,,\qquad \frac{\delta S}{\delta
       \beta}=\frac{1}{2}(d^2-r^2)=0\,,\phantom{\sum^1}
\end{equation}
where the following notation is introduced:
\begin{equation}\label{v_def}
      \phantom{\frac12} v\equiv\dot{x}-\dot{d\,\,}+\frac{s}{mr^2}[\,d,\dot{d\,\,}]\,.\phantom{\frac12}
\end{equation}
At first we note that the Lagrange multipliers $\alpha, \beta$ can
be algebraically expressed from equations of motion (\ref{eom1}),
(\ref{eom2}),
\begin{equation}\label{alpha_beta}
       \alpha=0\,,\qquad \beta=
       \frac{2s}{r^4}\frac{(\,d,v,\dot{d\,\,})}{\sqrt{-(v,v)}}\,.
\end{equation}
This observation means that the Lagrange multipliers $\alpha,\beta$
are auxiliary variables, while all the physical dynamics is
described by the variables $x,d$. Once the Lagrange multipliers are
expressed, the remaining independent equations include (\ref{eom1}),
and constant separations conditions (\ref{eom3}). From (\ref{eom1}),
it follows
\begin{equation}\label{v_dotv_vect}
       \phantom{\frac12}[\,v,\dot{v}\,]=0\,.\phantom{\frac12}
\end{equation}
Equation (\ref{dxddx}) is a consequence of the definition of vector
$v$ (\ref{v_def}), and relations (\ref{eom3}), (\ref{v_dotv_vect}).
This means that the equations (\ref{d-cyl}), (\ref{dxddx}) follow
from the action functional (\ref{S}).

Let us now discuss the interpretation of the parameters $m$, $s$
involved in the series of functionals (\ref{S}). The action
(\ref{S}) is Poincare-invariant. The infinitesimal rotations and
translations of Minkowski space induce a following transformations
of dynamic variables:
\begin{equation}\label{symmetry}
       \phantom{\frac12}\delta x=[\omega,x]+a\,, \qquad \delta d=[\omega,d]\,,\qquad \delta\alpha=0\,,\qquad\delta\beta=0\,,\phantom{\frac12}
\end{equation}
where $a,\omega$ are transformation parameters. The integrals of
motion, which correspond to this symmetry, are the momentum $p$ and
total angular momentum $J$,
\begin{equation}\label{pJ}
       \phantom{\frac12}p= \frac{mv}{\sqrt{-(v,v)}}\,,\qquad
       J=\frac{m[x-d,v]-sv}{\sqrt{-(v,v)}}\,.\phantom{\frac12}
\end{equation}
The mass-shell and spin-shell constraints (\ref{mass-spin-shell})
are satisfied by these quantities, as it is expected for the
irreducible spinning particle with mass $m$ and spin $s$.

It is interesting to compare the action (\ref{S}) for spinning and
spinless particles. Consider the value of the classical action
(\ref{S}) in the special class of paths with $n=(1,0,0)$ and
$a=(0,0,0)$. Modulo reparametrization, the general casual trajectory
on such a cylinder read
\begin{equation}\label{example}
       \phantom{\frac12}x(\tau)=(\kappa\tau,\,r\cos\psi(\tau),\, r\sin\psi(\tau) )\,,
       \qquad
       d(\tau)=(0,\,r\cos\psi(\tau),\,r\sin\psi(\tau) )\,. \phantom{\frac12}
\end{equation}
where $\kappa$ is a real constant having the sense of speed, and
$\psi(\tau)$ is some function. The substitution of this path into
action gives
\begin{equation}\label{}
    \phantom{\frac12}S(\tau_1,\tau_0)=-m\kappa(\tau_1-\tau_0)-s\left(\psi(\tau_1)-\psi(\tau_0)\right)\,.\phantom{\frac12}
\end{equation}
This expression is a sum of two terms. The first one is the length
of particle's path projection on the cylinder axis. The second one
is angle of rotation of particle's path around the cylinder. For
spinless particle the length of the particle path projection is the
only contribution in the action. Once spin is nonzero, the rotation
angle is also included in the action functional. In $d=4$ Minkowski
space, the similar dependence of the action on the mass and spin has
been observed in \cite{Bratek}.

\section{COMPARISON WITH PREVIOUS RESULTS}
\par
Let us show that the previously known model of spinning particle
from ref. \cite{GKL} can be derived from the action principle
(\ref{S0}). For reasons of simplicity, we consider only the case of
isotropic spin vector,
\begin{equation}\label{special_case_}
    r=\frac{s}{m}\,.
\end{equation}
In this setting, the constant separation conditions (\ref{d-cyl})
can be expressly solved by means of normalized isotropic vector
$\xi$,
\begin{equation}\label{def_dtild}
d=-\frac{s}{m}\frac{[n,\xi]}{(n,\xi)}\,,\qquad
n=\frac{\dot{x}}{\sqrt{-\dot{x}^2}}\,,\qquad
\xi\equiv(1,\,\sin\varphi,\,\cos\varphi)\,,
\end{equation}
where $\varphi$ is a new dynamical variable. Once the difference
vector $d$ is expressed in terms of $x,\varphi$, the action
functional (\ref{S}) takes the previously known form,
\begin{equation}\label{isotr}
    S[x(\tau),\varphi(\tau)]=-m\int\sqrt{-\dot{x}^2\bigg(1-2\frac{s}{m}\frac{\dot{\varphi}}{(\dot{x},\xi)}\bigg)}\,\,d\tau\,.
\end{equation}

\section{CONCLUSION}
In the present work, the variational model of irreducible massive
spinning particle is constructed from the requirement that the class
of gauge equivalence of its paths forms a cylindrical surface in
Minkowski space. The specifics of the problem is that both the
spinning and spinless particles have one and same classical paths,
while the difference between the models appears at the quantum
level. To introduce spin, we propose the series of action
functionals, which have one and the same Euler-Lagrange equations.
The different representatives in the action functional series are
not equivalent: the action of spinless particle is just the length
of the projection of path onto the cylinder axis, while in the
presence of spin the rotation of the trajectory around cylinder is
taken into account. To our knowledge, the constructed spinning
particle model is one of a few Lagrangian theories, in which
different variational formulations are physically relevant. In the
final part of the work, we showed that in the certain particular
case our model is equivalent to the previously known one.

\section{ACKNOWLEDGMENTS}
We thank A.A. Sharapov for discussions and valuable comments. IAR
expresses his gratitude to the organisers of the AYSS–-2019
conference for their hospitality and support. This research was
funded by state task of Ministry of Science and Higher Education of
Russian Federation, grant number 3.9594.2017/8.9.


\end{document}